# Classification of executive functioning performance post-longitudinal tDCS using functional connectivity and machine learning methods

Akash K Rao
*Applied Cognitive Science Laboratory*
*Indian Institute of Technology Mandi*
Mandi, India
0000-0003-4025-1042

Vishnu K Menon
*Applied Cognitive Science Laboratory*
*Indian Institute of Technology Mandi*
Mandi, India
0009-0007-9449-0934

Shashank Uttrani
*Applied Cognitive Science Laboratory*
*Indian Institute of Technology Mandi*
Mandi, India
0000-0003-2601-2125

Ayushman Dixit
*School of Computing and Electrical Engineering*
*Indian Institute of Technology Mandi*
Mandi, India
0000-0002-7733-0978

Dipanshu Verma
*School of Computing and Electrical Engineering*
*Indian Institute of Technology Mandi*
Mandi, India
0000-0003-2461-5547

Varun Dutt
*Applied Cognitive Science Laboratory*
*Indian Institute of Technology Mandi*
Mandi, India
0000-0002-2151-8314

*Abstract*— Executive functioning is a cognitive process that enables humans to plan, organize, and regulate their behavior in a goal-directed manner. Understanding and classifying the changes in executive functioning after longitudinal interventions (like transcranial direct current stimulation (tDCS)) has not been explored in the literature. This study employs functional connectivity and machine learning algorithms to classify executive functioning performance post-tDCS. Fifty subjects were divided into experimental and placebo control groups. EEG data was collected while subjects performed an executive functioning task on Day 1. The experimental group received tDCS during task training from Day 2 to Day 8, while the control group received sham tDCS. On Day 10, subjects repeated the tasks specified on Day 1. Different functional connectivity metrics were extracted from EEG data and eventually used for classifying executive functioning performance using different machine learning algorithms. Results revealed that a novel combination of partial directed coherence and multi-layer perceptron (along with recursive feature elimination) resulted in a high classification accuracy of 95.44%. We discuss the implications of our results in developing real-time neurofeedback systems for assessing and enhancing executive functioning performance post-tDCS administration.

I. INTRODUCTION

Executive functioning is a set of higher-order cognitive processes necessary for goal-directed activities, decision-making, problem-solving, and adaptive functioning in everyday life [1]. It entails various mental processes that assist individuals in organizing, planning, initiating, and monitoring their actions and managing their emotions and attention [1]. The prefrontal cortex (PFC), a region of the brain positioned directly behind the forehead, is the center of executive functioning [1]. The dorsolateral prefrontal cortex (DLPFC) is a significant subregion important in working memory, cognitive flexibility, and attentional regulation. Another important component of the executive functioning network is the anterior cingulate cortex (ACC), which is responsible for conflict resolution, error detection, and decision-making [1]. Moreover, the executive functioning system has an elevated level of interaction with different brain regions engaged in diverse cognitive tasks.

In recent years, there has been growing interest in evaluating the efficacy of different techniques to enhance executive functioning [2]. Among these techniques, non-invasive brain stimulation techniques, like tDCS, have found considerable traction in the past decade. tDCS modifies neuronal activity by passing weak electrical currents through specific brain areas. tDCS applied across the dorsolateral prefrontal cortex (DLPFC), a critical brain region involved in executive functions, has been found in studies to improve working memory performance [2,3]. Anodal stimulation of the DLPFC is hypothesized to increase excitability, resulting in a greater capacity for working memory and cognitive flexibility. Long-term potentiation (LTP), a process related to learning and memory, is thought to be enhanced by anodal tDCS stimulation. Still, cathodal tDCS stimulation has been shown to promote long-term depression (LTD) by influencing synaptic connections and neural networks engaged in executive tasks [2]. However, an empirical evaluation of how longitudinal tDCS, i.e., tDCS administration over time, affects the neurophysiological basis of executive functioning and subsequent decision-making prowess is mainly missing and much needed in the literature.

In recent years, researchers have successfully combined Electroencephalography (EEG) data with machine learning/deep learning techniques to gain valuable insights into brain function and human information processing and develop predictive models for cognitive performance [5]. EEG is a non-invasive neuroimaging technique that analyses brain electrical activity using electrodes on the scalp. Spectral power, event-related potentials (ERPs), connection measurements, and other time-domain or frequency-domain metrics are the frequently employed techniques in EEG analysis [5]. Source-space-based functional connectivity analysis for EEG has gained considerable interest in recent years [5,6]. Functional connectivity refers to the temporal correlation and synchronization of neural activity between brain regions, providing insights into the interactions and networks underlying cognitive processes, behavior, and various neurological conditions [4].

Machine learning methods have been shown to have the advantage of discovering discrete correlations and non-linear dynamics in EEG data that traditional statistical approaches might overlook [9]. In the recent past, machine learning techniques like support vector machines (SVM), gradient boosting machines (GBMs), and random forest (RF) have been used to classify lower-order cognitive processes like workload, working memory, etc., using generic EEG features like ERPs and spectral power [8]. However, the efficacy of combining functional connectivity analysis and machine learning techniques to predict executive functioning performance, especially after tDCS administration, is missing and much needed in the literature.

We seek to fill this gap in the research by using EEG-based functional connectivity and machine learning models to classify executive functioning performance in a standard psychometric task following tDCS administration over time. The work is significant since machine learning algorithms based on functional connectivity analyses were created to predict executive functioning ability. The following section briefly overviews prior research on predicting cognitive states and performance using EEG and machine learning technologies. The experiment and the numerous functional connectivity studies and machine learning models are then encapsulated. We conclude by analysing our findings and emphasizing the importance of accurately forecasting changes in executive functioning performance following tDCS administration.

## II. BACKGROUND

In recent years, EEG and machine learning techniques have been employed to predict generic cognitive processes like workload, attention, and working memory [17]. A wide range of machine learning methods, including artificial neural networks (ANN), support vector machine (SVM), and random forest, have been used to discriminate mental workload levels using EEG patterns for binary (low- vs. high-workload levels) or multilevel workload classification [5]. Nonetheless, most of this research has relied on generic EEG features, resulting in low classification accuracies and lower generalizability and reproducibility across complicated tasks [6-8]. The significant variability of cognitive processes usually makes accurate classification across different cognitive activities, subjects, and sessions difficult. Recently, state-of-the-art classifiers, such as the hierarchical Bayes model and adaptive SVM, have significantly addressed higher-order workload distinction in complicated tasks (such as multitasking paradigms), attaining high classification accuracy [9]. However, executive functioning performance classification remains difficult, as earlier attempts have usually yielded poor results. For example, researchers in [6] used ANN to achieve high binary classification accuracy of working memory performance, but the subsequent multi-class classification yielded much lower accuracy (less than chance) [6]. After their insufficient cross-task classification accuracy, researchers in [7] effectively predicted multi-class workload on two working memory (WM) tasks using regression. However, assessment by mean square error in regression is insufficient to compare with classification methods accurately [7].

Recent studies that use network modeling to represent the complex interactions among brain areas lend an improved understanding of the brain's functional structure by obtaining attributes such as dynamic rewiring of functional connections among brain areas [4]. Increasing research combining classification approaches with brain networks has demonstrated the ability to find discriminatory network connections between different situations in human cognition [9]. Researchers in [10] examined the cross-frequency functional connectivity metric in a mental arithmetic task and discovered mental workload-related interactions between frontal theta and parieto-occipital higher alpha frequencies. Similarly, researchers in [9] implemented CNN-based classification based on functional connectivity using fMRI data to distinguish between distinct cognitive states. However, the prediction of executive functioning performance (a higher-order cognitive process) using functional connectivity metrics and machine learning, especially after tDCS intervention, is lacking and much needed in literature. In our research, we have addressed this literature gap by employing three functional connectivity metrics and four machine learning algorithms to classify executive functioning performance post-longitudinal tDCS administration.

## III. METHODS

### A. Experiment design

Fifty participants (31 males, 19 females, mean age = 22.56 years, SD = 2.78 years) from the Indian Institute of Technology Mandi, Himachal Pradesh, India participated in the experiment. The Institutional ethical committee approved the study, and the experiment was conducted in strict compliance with the declaration of Helsinki. The experiment consisted of three phases: pre-intervention (Day 1), intervention (Day 2 to Day 8), and post-intervention (Day 10). All participants completed the pre-intervention phase on Day 1. They were informed about the experiment and its objectives and assessed for potential risks during tDCS administration using a screening questionnaire []. After obtaining written consent, demographic data was collected, and EEG data acquisition was conducted using a 32-channel system. The participants eventually executed the permuted rules operation task for assessing executive functioning performance with simultaneous EEG data was recorded. Participants were then randomly divided into two conditions: experimental and placebo control. The experimental group received 2mA anodal tDCS intervention for 15 minutes on the dorsolateral prefrontal cortex (dlPFC) every day from Day 2 to Day 8 during simultaneous task training. A pre- and post-tDCS intervention questionnaire evaluated potential physical effects [2]. The control group received sham tDCS intervention (i.e., electrodes placed but no tDCS applied) during the same period. After a 24-hour break, participants executed the permuted rules operations task again (with simultaneous EEG being acquired) on Day 10. We utilized the Caputron Activadose 2 tDCS device to administer anodal tDCS, with the anode placed on the left dorsolateral prefrontal cortex (dlPFC).

### B. Permuted Rules Operations Task (PROT)

The Permuted Rules Operation Task (PROT) is a psychometric task used in psychological and neuropsychological studies to test cognitive flexibility and executive functioning [11]. The task was designed using OpenSesame 3.9.2. It is intended to assess an individual's propensity to switch between rules and respond to changing task demands [11]. It entails a series of trials in which participants are provided with stimuli (in this experiment, words) and are instructed to respond following certain rules. The test's major characteristic is that the rules change regularly and unpredictably during the task, prompting individuals to alter their cognitive strategies accordingly [11].

For instance, in one trial, participants are presented with two words (e.g., "pillow" and "rock"). The participants are instructed to respond to the two words as quickly as possible according to three different rules presented beforehand. The three rules presented beforehand included:

a) Semantic sensory rules (Such as "Is it BROWN?", "Is it SOFT?", "Is it SWEET?", "Is it LOUD?")

b) Logical rules, where the participants were asked to evaluate the words according to semantic rules stated in (a) and following sub rules: "Is your answer the same for both words?", "Is your answer different for both words?", "Is your answer true for the first word?", "is your answer false for the first word?". If the answer to the relevant question was yes, the participant was asked to press the current "yes" key as specified in (c)

c) Rules for appropriate motor response, where the participants were provided with appropriate "yes" or "no" keys to press following the rules presented in (a) and (b).

Therefore, if the rules provided to evaluate the pair "pillow" and "rock" is "Is it SOFT?" and "Is your answer the same for both words?", then the participants had to respond with the corresponding "no" key because both "pillow" and "rock" are not soft.

The trial phase consisted of 8 trials, where feedback was provided after each trial.

Funding agency: Life Sciences Research Board, Defence Research and Development Organization (IITM/DRDO-LSRB/VD/301).

The test phase consisted of 72 trials (2 blocks of 36 trials each). A one-minute break was provided between each block in the test phase. The stimuli (i.e., "pillow", "rock," etc.) was selected based on the normalization study conducted by []. The interstimulus interval was kept to 1500ms. Performance parameters like percentage accuracy and mean latency (in milliseconds). The percentage accuracy was calculated by:

$$Percentage\ Accuracy = \frac{Number\ of\ correct\ responses}{72\ (Number\ of\ trials)} * 100 \quad (1)$$

As elucidated below, the percentage accuracy was taken as the output variable to be predicted post-tDCS using different functional connectivity and machine learning algorithms. The difference in the percentage accuracy on Day 10 and Day 1 was calculated and used as the basis for eventual classification. The difference in percentage accuracy was found to be normally distributed; based on this condition, a three-class classification paradigm was designed based on the obtained mean (20.76) and standard deviation (14.78). The first class was labeled as low cognitive enhancement (range $\mu$ to $\mu+\sigma$ = 20.76 to 35.54), the second class as medium cognitive enhancement (range $\mu+\sigma$ to $\mu+2\sigma$ = 35.54 to 50.32), and the third class as high cognitive enhancement (range $\mu+2\sigma$ to $\mu+3\sigma$ = 50.32 to 65.1).

### C. EEG data acquisition and analysis

The EEG data was collected using 32 Ag/AgCl saline sensor electrodes placed using the conventional 10-20 electrode placement system (Emotiv EPOC Flex and EMOTIVPRO data collecting software, v2.34b, San Francisco, USA). The EEG data was collected at a sampling rate of 256 Hz. Throughout the session, the electrode impedance was kept below 10 K$\Omega$. Anti-aliasing (0.1-45 Hz) with a band-pass filter and a 50 Hz notch filter was used to eliminate all major interferences. Before the start of the experiment, baseline EEG data was collected (for 60 seconds), where the subjects were asked to relax and keep their eyes open. Brainstorm [12] was used for pre-processing and feature extraction from EEG data. It is documented as a MATLAB plugin and is freely available for distribution under the GNU public license [12]. The raw EEG data was initially band-pass filtered from 0.1 to 45 Hz. The band-pass filtered data was then re-referenced to the average of the electrodes in the left and right mastoid. Picard's Independent Component Analysis (ICA) technique detected and removed eye blink artifacts [12]. The artifact-rejected EEG signal was then modified with respect to the baseline data.

### D. Source Localization

We designed a forward model of the 32 electrodes in the sensor place for source localization using the symmetric boundary element method [13]. We then used dynamic statistical parametric mapping (dSPM) in the frequency domain to project the sensor-space, temporal EEG signal into the sources of the brain. We used a standard ICBM 152 T1 image from the Brainstorm plugin [13]. Using the default tissue conductivity values, a volume conduction model was developed. Electrode placements on the scalp were mapped, and sources were limited to grey matter. Using a Hanning window, the power spectrum and cross-spectrum between EEG channels were recovered for the beta frequency band (13-29 Hz). Multi-taper analysis was used to localize the source, resulting in activation images with 7556 voxels at 5mm spatial resolution [13].

The current spectral density obtained was further parcellated into 28 different regions of interest (ROIs; 14 in each hemisphere) according to the Brodmann atlas [14]. The ROIs included the primary somatosensory area, primary motor areas (anterior/posterior), pre-motor area, Broca's area (pars opercularis and pars triangularis), primary/secondary visual area, the visual area in the middle temporal lobe, and the entorhinal and perirhinal cortex [14].

*E. Functional connectivity metrics*

*1) Magnitude squared coherence (MSC)*

MSC is a functional connectivity metric used in neuroscience to evaluate the degree of synchronization across various brain areas [14]. It calculates the coherence squared, which reflects the proportion of their spectral power that is jointly shared, to measure the frequency-domain connection between two signals [14].

The MSC of the signal A and B is given as:

$$MSC_{AB}(f) = \frac{PSD_{AB}(f)^2}{|PSD_{AA}(f)| \times |PSD_{BB}(f)|} \quad (2)$$

where $PSD_{AA}(f)$ and $PSD_{BB}(f)$ are the power spectral densities of the signals A and B, respectively, and $PSD_{AB}(f)$ is the cross-power spectral density at frequency $f$.

*2) Wavelet coherence (WC)*

Wavelet coherence analyzes the temporal evolution of two signals over different frequency components, making identifying areas of strong synchronization or coupling easier [14]. In contrast to classic coherence approaches that rely on the Fourier transform, wavelet coherence provides a time-resolved and localized approach appropriate for capturing non-stationary and transient interactions [14].

The wavelet transform of a signal $a$ is a function of both time and frequency [14]. It is given as the convolution of the input with a wavelet family $\varphi_u$:

$$W_a(t,f) = \int_{-\infty}^{\infty} a(u) \cdot \varphi_{t,f}^*(u) du \quad (3)$$

The cross-spectrum wavelet around time $t$ and frequency $f$ (given as input signals $a$ and $b$) are derived from the wavelet transforms of $b$ and $c$:

$$CSW_{AB}(t,f) = \int_{t-\frac{\delta}{2}}^{t+\frac{\delta}{2}} W_a(\tau,f) \cdot W_b^*(\tau,f) d\tau \quad (4)$$

where * represents the complex conjugate and $\delta$ is a frequency-dependent scalar. $\tau$ is the wavelet coefficient at time $t$.

Therefore, the wavelet coherence at time $t$ and frequency $f$ is given as

$$WC_{AB}(t,f) = \frac{|CSW_{AB}(t,f)|}{|CSW_{AA}(t,f) \times CSW_{BB}(t,f)|^{\frac{1}{2}}} \quad (5)$$

*3) Partial directed coherence (PDC)*

PDC emanates from multivariate autoregressive modeling, a statistical approach in which each time series is modeled as a linear combination of its previous and past values of other time series [15]. It enables the discovery of causal relationships between the system's components by calculating the model's coefficients [15]. The purpose of PDC is to capture the "directedness" of these interactions, determining the strength and directionality of the flow of information between network nodes [15].

Consider $A(t)$ is a set of signals estimated from $N$ EEG electrodes recorded:

$$A = [a_1(t), a_2(t), \ldots \ldots a_N(t)]^T \quad (6)$$

The multivariate autoregressive modeling process is an expressive description of the dataset $A$:

$$\sum_{r=0}^{p} B(r)A(t-r) = W(t) \quad (7)$$

where $W(t)$ is a zero-mean uncorrelated, multivariate white noise vector. $B(r)$ is the matrix of the autoregressive coefficients, and its elements show the influence of $A_j(t-r)$ on $A_i(t)$, and $p$ represents the model border.

The PDC is given as follows:

$$\varepsilon_{ij}(f) = \frac{\bar{B}_{ij}(f)}{\sqrt{\sum_{m=1}^{N} \bar{B}_{mj}(f)\bar{B}_{mj}^*(f)}} \quad (8)$$

where $B_{ij}(f)$ is the frequency domain representation of $a_{ij}(r)$

$$\bar{A}_{ij}(f) = \begin{cases} 1 - \sum_{r=1}^{p} a_{ij}(r)e^{-j2\pi fr} & , if\ i = j \\ -\sum_{r=1}^{p} a_{ij}(r)e^{-j2\pi fr} & , otherwise \end{cases} \quad (9)$$

We derived 28*28 adjacency matrices for each of the 50 participants (both for pre- and post-tDCS intervention) for the three functional connectivity metrics (MSC, WC, PDC). These adjacency matrices were given as the input to the different machine learning algorithms. We also employed two different feature selection techniques to systematically explore the dataset (comprising of 784 features for each subject from each of the three functional connectivity metrics) for the most relevant and informative subset of features. Out of 784 features for each subject, the top 100 features were selected for further analysis for each functional connectivity metric. The feature selection techniques employed in this research work were:

*a) Forward Feature selection (FFS):* It is a variant of the sequential feature selector (SFS), begins with an empty set and adds one feature at a time, choosing the one which results in the greatest improvement in model performance [16]. This procedure is repeated until a predetermined amount of features are chosen, or the model's performance no longer increases appreciably [16].

*b) Recursive feature elimination (RFE):* The RFE method begins with training a model on all the features and assigns a weight or priority score to each feature depending on its contribution to the model's performance [16]. The features with the lowest weights or significance ratings is then pruned [16]. The model is retrained on the decreased feature set, and the feature reduction and model retraining procedure is repeated until a predetermined number of attributes or a stopping criterion is met [16].

*F. Machine learning algorithms*

We employed different machine learning algorithms for classifying executive functioning performance based on the functional connectivity algorithms explained in subsection F. The different machine learning algorithms used are as given below:

*a) Support Vector Machine (SVM):* SVM operates by determining the appropriate hyperplane for separating various classes in the data space [17]. The hyperplane is chosen to maximize the margin, which is the distance between the hyperplane and the nearest data points of each class, also known as support vectors [17].

*b) Decision Trees (DT):* A decision tree is built by recursively splitting data based on feature values to build subsets that are as pure as feasible, which means that each subset mostly comprises instances of a single class [17].

*c) Random Forest (RF):* The Random Forest approach generates many decision trees during training by selecting random subsets of the original dataset and random subsets of characteristics for each tree [17]. Each decision tree in the Random Forest is built using a technique known as recursive partitioning, which involves repeatedly splitting the data into subsets depending on the most discriminatory attributes, resulting in a tree-like structure [17].

*d) Multi-layer perceptron (MLP):* The MLP is made up of numerous layers of linked neurons that are placed sequentially [17]. The layers are organized into three sections: an input layer that accepts input data, one or more hidden layers that do intermediary computations, and an output layer that produces final predictions or outputs [17].

The different variations in the hyperparameters used in various machine learning algorithms are shown in Table 1. To pick the optimal parameters for the machine learning model, we utilized 10-fold cross-validation (three times) [17]. All of the machine learning models were built in Python using scikit-learn. The machine learning models were trained for several hyperparameters, and the hyperparameters with the highest test accuracy during training folds were deemed the best. We employed grid search to determine the best set of hyperparameters for each machine learning model. The machine learning models were trained for the various hyperparameters, and the hyperparameters with the highest accuracy were considered the best. The symmetric difference between the functional connectivity measures (MSC, WC, PDC) on Day 1 and Day 10 was used as input to the machine learning models, and the difference in the percentage accuracy on Day 1 and Day 10 was taken as the output variable to be classified.

TABLE I. DIFFERENT HYPERPARAMETERS AND THE CORRESPONDING RANGE OF VALUES USED IN DIFFERENT MACHINE LEARNING ALGORITHMS

| Machine learning model | Hyperparameters varied |
|---|---|
| Support Vector Machine | 1) C – 0.01 to 100 (steps of 0.01)<br>2) γ – 0.001 to 1 (steps of 0.001)<br>3) Kernel – Linear, polynomial, radial basis function |
| Decision Tree | 1) Maximum depth – 2 to 10 (steps of 1)<br>2) Minimum samples split – 2 to 10 (steps of 1)<br>3) Minimum samples leaf – 1 to 10 (steps of 1) |
| Random Forest | 1) Maximum depth – 2 to 10 (steps of 1)<br>2) Minimum samples split – 2 to 10 (steps of 1)<br>3) Minimum samples leaf – 1 to 10 (steps of 1)<br>4) Number of estimators – 10 to 100 (steps of 10) |
| Multi-layer perceptron | 1) Hidden layer sizes – 1 to 3 (steps of 1)<br>2) Hidden nodes count – 10 to 1000 (steps of 10)<br>3) Activation function – Logistic, tanh, rectified linear unit<br>4) Solver – Adam, stochastic gradient descent<br>5) Alpha - 0.0001 to 0.1 (steps of 0.001) |

IV. RESULTS

Table 2 shows the cross-validation test accuracies (in percentage) obtained for different functional connectivity algorithms combined with different feature selection techniques and machine learning algorithms.

TABLE II. TEST ACCURACY FOR THE DIFFERENCE IN THE PERCENTAGE ACCURACY BETWEEN DAY 1 AND DAY 10 OF THE PROT TASK FOR DIFFERENT FUNCTIONAL CONNECTIVITY METRICS, FEATURE SELECTION TECHNIQUES, AND MACHINE LEARNING ALGORITHMS

| Functional connectivity metric | Feature selection technique | SVM | DT | RF | MLP |
|---|---|---|---|---|---|
| MSC | FFS | 56.78% | 59.44% | 53.45% | 62.33% |
|  | RFE | 52.56% | 58.78% | 62.81% | 69.78% |
| WC | FFS | 69.45% | 75.55% | 72.33% | 80.33% |
|  | RFE | 67.55% | 78.55% | 83.55% | 85.55% |
| PDC | FFS | 63.44% | 79.33% | 84.77% | 89.55% |
|  | RFE | 78.55% | 83.44% | 89.55% | **95.44%** |

Table 3 shows the best set of hyperparameters obtained during model calibration.

TABLE III. BEST SET OF HYPERPARAMETERS FOR DIFFERENT MACHINE LEARNING ALGORITHMS OBTAINED DURING MODEL CALIBRATION

| Machine learning model | Optimal hyperparameters during model calibration |
|---|---|
| SVM | - Kernel – linear, γ = 0.001, C = 1 |
| DT | - Minimum samples split = 4, Minimum samples leaf = 8, Maximum depth = 2 |
| RF | - Number of estimators = 40, Minimum samples split = 5, Minimum samples leaf = 4, Maximum depth = 9 |
| MLP | - Solver – adam, hidden layer sizes = 50, α = 0.1, activation = reLu |

As shown in Table 2, we obtained the highest three-class classification accuracy of 95.44% with the MLP, combined with PDC as the functional connectivity metric and RFE as the feature selection technique.

## V. DISCUSSION AND CONCLUSION

This research aimed to efficiently classify executive functioning performance using functional connectivity analysis, feature selection techniques, and machine learning algorithms. Results revealed that PDC-RFE-MLP model yielded a high classification accuracy of 95.44%. This study used a source localization technique (dSPM) to extract functional connectivity metrics compared to previous works. The significantly better classification results compared to other works [6-8] were consistent with [10]. They reasoned that due to the excellent spatial resolution and reduced volume conduction offered by dSPM (through the incorporation of boundary element methods and anatomical information), source estimation provided a more accurate and detailed representation of the underlying neural processes [10]. PDC yielded a higher classification accuracy compared to other functional connectivity metrics. These results were consistent with [15], who elucidated that PDC was able to extract significantly more information from the brain's functional networks. PDC inherently provides directional information, indicating which brain regions causally influence the others. Researchers in [15] also reasoned that since PDC is rooted in Granger causality [15], it mitigates volume conduction more efficiently than other metrics employed in this study. RFE and MLP well exploited these inherent advantages of using PDC. Researchers in [16] had reasoned that RFE's propensity to handle multicollinearity and offer model agnosticism was preferred compared to other feature selection techniques for drowsiness detection [16]. This combined with MLP's capability to model complex and non-linear relationships, led to a higher classification accuracy in executive functioning performance than the predecessors. However, this research work is not devoid of limitations. Even though the source localization technique employed a cortical parcellation technique (dSPM), it is still prone to volume conduction effects due to the inherent limitations in the EEG hardware. This disadvantage could be mitigated by employing an EEG acquisition system with more channels (>64) in the future. In addition, in the future, we intend to employ bleeding-edge deep neural networks that can interpret the spatial-temporal relationship between brain networks more efficiently. This might lead to a better understanding of the underlying cortical dynamics during executive functioning. This framework can potentially be used to design cognitive state assessors for real-time executive functioning performance prediction post-tDCS administration.